\documentclass[10pt,twocolumn,letterpaper]{article}

\usepackage{cvpr}
\usepackage{times}
\usepackage{epsfig}
\usepackage{graphicx}
\usepackage{amsmath}
\usepackage{amssymb}
\usepackage{booktabs}

\usepackage{graphicx}
\graphicspath{{figures/}}

\usepackage[pagebackref=true,breaklinks=true,letterpaper=true,colorlinks,linkcolor =blue,urlcolor = black,bookmarks=false]{hyperref}

\cvprfinalcopy 

\def\cvprPaperID{****} 

\ifcvprfinal\fi
\begin{document}


\title{Graph Neural Network for Product Recommendation on the Amazon Co-purchase Graph}

\author{
Mengyang Cao\thanks{All authors contributed equally to this work.},
Frank F. Yang\footnotemark[1],
Yi Jin\footnotemark[1],
Yijun Yan\footnotemark[1] \\
Department of Computer Science, Georgia Institute of Technology, Atlanta, GA, US \\
{\tt\small \{mcao45, frank.yang, yjin432, yyan389\}@gatech.edu}
}

\maketitle

\begin{abstract}
   Identifying relevant information among massive volumes of data is a challenge for modern recommendation systems. Graph Neural Networks (GNNs) have demonstrated significant potential by utilizing structural and semantic relationships through graph-based learning. This study assessed the abilities of four GNN architectures, LightGCN, GraphSAGE, GAT, and PinSAGE, on the Amazon Product Co-purchase Network under link prediction settings. We examined practical trade-offs between architectures, model performance, scalability, training complexity and generalization. The outcomes demonstrated each model's performance characteristics for deploying GNN in real-world recommendation scenarios.

\end{abstract}

\section{Introduction}
Recommender Systems (RecSys) are capstone of modern e-commerce platforms like Amazon, Shopify, Walmart and Ali Group Holding, helping users discover relevant products and lead to significant revenue growth through personalized suggestions. Conventional recommendation systems on sites like Amazon rely heavily on matrix factorization techniques \cite{koren2009matrix} like SVD and collaborative filtering strategies like item based or user based nearest neighbors. Although co-occurrence signals from previous user-item interactions have been effectively collected by these methods, they frequently fall short in their capacity to take into account higher order structural information present in the product graph. Additionally, they often make the assumption that relationships are static and have trouble generalizing to new or sparsely connected products, which results in less than ideal performance in cold start situations, and limiting their generalization in dynamic settings. 

Recent work has focused on shallow graph models like Node2Vec \cite{grover2016node2vec} and DeepWalk\cite{perozzi2014deepwalk}. use They learns node embeddings by using random walks in a graph. Although computationally efficient, they tend to overlook important details like node and edge features, assume uniform graph structures, and do not emphasize more informative neighbors due to the lack of "attention system". 

Moreover, these previous methods are less scalable and less appropriate for real-time recommendation tasks since they frequently call for complete retraining to integrate new nodes or interactions.On the other hand, graph neural networks (GNNs) like GraphSAGE and GAT can combine structural and feature information. They also provide inductive capabilities that let them generalize to nodes that are not visible. Still, GNNs are not widely used in production systems yet because they are expensive to compute, hard to train on large graphs in small batches, and hard to tune parameters across different types of products and interactions \cite{wu2020comprehensive}.

Graph Neural Networks (GNNs) offer a more expressive framework by jointly leveraging graph structure and
node features. Architectures such as GraphSAGE \cite{hamilton2017graphsage} perform inductive neighborhood aggregation to generalize unseen nodes. GAT use attention mechanisms\cite{veličković2018graphattentionnetworks} to adapt weight neighbor influence. PinSAGE\cite{ying2018pinsage} introduces random walk based neighbor sampling to billion-node graph for webscale recommendation. LightGCN \cite{he2020lightgcn} simplifies graph convolution for collaborative filtering by removing features transformations and nonlinearity. In general, they provide a comprehensive basis for understanding the trade-offs between model complexity, performance, and scalability in real-world recommendation systems. A high-level comparison is shown in Table~\ref{tab:model_intro_comparison}.

\begin{table*}[t]
\centering
\scriptsize  
\begin{tabular}{lccccc}
\toprule
\textbf{Model} & \textbf{Designed For} & \textbf{Uses Features} & \textbf{Aggregation} & \textbf{Adaptive Neighbor Weighting} & \textbf{Scalable} \\
\midrule
LightGCN  & Collaborative filtering (CF) & No  & Neighborhood sum (no transform) & No  & Yes \\
GraphSAGE & General link prediction       & Yes & Mean pooling                    & No  & Yes \\
GAT       & Node classification           & Yes & Attention mechanism             & Yes & Partial \\
PinSAGE   & Web-scale recommendation      & Yes & Importance-weighted sampling    & Yes & Yes (web-scale) \\
\bottomrule
\end{tabular}
\caption{Comparison of GNN architectures by core design characteristics.}
\label{tab:model_intro_comparison}
\end{table*}

While these four models form the basis of our evaluation, each comes with known limitations in scalability, efficiency, or adaptability to certain recommendation scenarios. Addressing such limitations has been an active area of research, with several complementary techniques proposed in the literature. Scalability and efficiency have been improved through graph coarsening methods such as CONVMATCH~\cite{convmatch2023}, which reduce graph size while preserving predictive power, and asymmetric modeling approaches such as AML~\cite{aml2023}, which use different encoders for head and tail nodes to lower computation cost.Mixed-precision quantization techniques like MixQ-GNN~\cite{mixqgnn2025} and distributed frameworks such as Roc~\cite{roc2020} or communication-efficient strategies like SpLPG~\cite{splpg2025} further accelerate GNN training on large-scale graphs.

Recommendation research has also expanded to handle richer inputs and evolving graphs. DAEMON~\cite{daemon2022} fuses textual, visual, and graph features to improve cold-start performance and mitigate selection bias, while temporal graph models such as GraphPro~\cite{graphpro2023} adapt to changing graph structures without full retraining. These approaches demonstrate promising pathways for making GNN-based recommendation more robust, efficient, and adaptable.

In this work, we benchmark four representative GNN architectures—LightGCN, GraphSAGE, GAT, and PinSAGE—on the Amazon Product Co-purchase Network \cite{snapamazon} under an inductive link prediction setup. This dataset includes a large collection of Amazon products and their co-purchase relationships, indicating which products are frequently purchased together. To complement these model comparisons, our work combines several underexplored aspects that offer a more realistic and practical perspective on GNN-based product recommendation. Specifically,  the inductive evaluation strategy  was implemented to avoid data leakage and better simulate cold-start scenarios. We also construct a rich and heterogeneous node embedding pipeline that integrates SBERT-based title representations, one-hot group labels, standardized numerical features, and dense category paths. . In addition, our evaluation emphasizes on runtime efficiency and scalability, highlighting trade-offs that are often overlooked in more controlled or idealized settings. While the current pipeline does not incorporate coarsening, asymmetric modeling, multi-modal fusion, or temporal adaptation, it establishes a strong baseline and identifies concrete directions for future improvements in large-scale, feature-rich, and dynamic recommendation environments.

From a research perspective, this work also sheds light on the practical trade-offs between different GNN models for large-scale recommendation tasks, potentially informing the design of future production grade systems. Ultimately, the success of this project could push the boundaries of real-time, graph-based recommendation in dynamic, data rich environments like e-commerce.

\section{Methodology}
\subsection*{2.1 Data Overview}
The \textbf{Amazon Product Co-purchase Network}\footnote{\url{https://snap.stanford.edu/data/amazon-meta.html}} was selected for this study, which is a product graph with nodes representing products that are sold on Amazon and edges indicating co-purchase relationships among products. This dataset offers a rich structural graph and related product features, but does not contain any user-specific data.

\begin{itemize}
  \item \textbf{Nodes:} 548,552 unique items.
  \item \textbf{Edges:} 1,788,725 undirected co-purchase links.
  \item \textbf{Graph structure:} Unweighted and undirected product–product graph.
  \item \textbf{Node features:} Metadata such as product name, category, and reviews.
\end{itemize}
\begin{table}[h]
\centering
\begin{tabular}{l r r}
\toprule
\textbf{Dataset} & \textbf{\#Rows} & \textbf{\#Unique Items} \\
\midrule
Item.csv     & 548,552   & 548,552 \\
Category.csv & 2,509,699 & 519,781 \\
Review.csv   & 7,593,244 & 402,724 \\
\bottomrule
\end{tabular}
\caption{Dataset breakdown by file and number of unique items.}
\label{tab:dataset-breakdown}
\end{table}

The analysis was restricted to the graph’s connected components. Negative samples for supervised link prediction were generated by randomly selecting and connecting product pairs without existing links. The resulting training and validation sets incorporated graph structure, node features, and binary edge labels.
\subsection*{2.2 Data Preparation}
\textbf{Item.csv:} 

Dropped products that are marked as discontinued, which account for approximately 1\% of the original data amount. These items lack meaningful features, therefore should be excluded from downstream modeling process. Additionally, applied logarithmic transformation on numerical columns which exhibit strong right skewness to reduce the influence of extreme outliers while preserving relative magnitudes.

\vspace{0.5em}
\textbf{Category.csv:} 

The first 4 levels in each product's category path was kept to reduce sparsity while retaining hierarchical information, . For items associated with multiple category paths, only selected the most representative one based on frequency to ensure that each product appears only once in the final category mapping.

\vspace{0.5em}
\textbf{Review.csv:} 

For products with multiple reviews,  aggregated the numerical columns by computing the average. This guarantees that each item appears only once in the review dataset, enabling efficient merging and modeling.

\vspace{0.5em}
\textbf{Final Merge:} 

Constructed the final item-level dataset by inner joining the cleaned versions of item, category and review datasets. During this process, also updated the \texttt{similarity\_list} column to remove references to products that were dropped in earlier steps. The resulting dataset was saved for modeling.

\subsection*{2.3 Graph Construction}

\textbf{Positive Edge Construction:} 
\noindent
Constructed a graph by parsing the \texttt{similarity\_list} column from the merged dataset. Each entry in this column encodes positive links between products based on behavioral or semantic similarity. These 0.8 million links form the positive edge set in the graph.

\vspace{0.5em}
\textbf{Negative Edge Sampling:} 

To reflect real world sparsity in training, randomly sampled negative edges from the complement set of the graph and implemented a negative sampling function that allows dynamic control over the negative to positive ratio. 

This adaptability enables future experimentation with different positive-negative ratios (e.g., 1:5 or 1:10), which is important for our link prediction task. It prevents the possible negative effects from class imbalance. In addition, including more negative samples encourages the model to better learn the decision boundary for unlinked nodes.

\vspace{0.5em}
\textbf{Label Generation:} 

Each edge between products is assigned a binary label (1 for positive edges, 0 for negative edges). These labels are used during model training for loss computation.

\subsection*{2.4 Feature embedding}
We generated node level embeddings for the Amazon product co-purchase graph by incorporating structured metadata, including title, category, category path, review data, and sales rank. To convert this heterogeneous information into structured node features suitable for input to graph neural networks, we apply a combination of classical and adjusted embedding strategies. The final node embedding is constructed by concatenating individual feature embeddings for each node.\\

\noindent
\textbf{Title Embedding} \\
Each product includes a descriptive title that often contains its semantic domain and intended use. For instance, titles like \textit{``Patterns of Preaching: A Sermon Sampler Book''} or \textit{``Catholic Bioethics and the Gift of Human Life''} contain critical domain related information. To embed these text descriptions, we employ the Sentence-BERT (SBERT) model\cite{reimers2019sbert}.

SBERT generates a dense vector representation for each title by encoding the full sentence into a fixed length embedding. For a title $t_i$ corresponding to product $i$, we obtain:

\[
\mathbf{x}_i^{\text{title}} = \mathrm{SBERT}(t_i),
\]

where $\mathbf{x}_i^{\text{title}} \in \mathbb{R}^d$ and $d$ is the embedding dimension. \\

\noindent
\textbf{Group Encoding} \\
Each product is also labeled with a high-level \texttt{group} label, indicating its domain, such as \texttt{book}, \texttt{music}, or \texttt{dvd}. This categorical information is encoded using one-hot encoding. \cite{sklearn_onehot}\\

\noindent
\textbf{Numeric Feature Embedding} \\
To further enrich the node representation, we extracted several numerical features that describe product popularity, review activity, and user engagement. The following fields are:

\begin{itemize}
    \item \texttt{salesrank\_log}
    \item \texttt{category\_count}
    \item \texttt{reviews\_total\_log}
    \item \texttt{reviews\_downloaded\_log}
    \item \texttt{reviews\_avg\_ratings}
    \item \texttt{reviews\_avg\_votes}
    \item \texttt{reviews\_avg\_helpful}
\end{itemize}

Each of these features is normalized using z-score standardization to ensure that all numeric dimensions are normalized and comparable\cite{james2013isl}:

\[
x_i^{(j)} = \frac{x_i^{(j)} - \mu_j}{\sigma_j},
\]

where $x_i^{(j)}$ is the $j$-th numeric feature of product $i$, and $\mu_j$, $\sigma_j$ are the mean and standard deviation of feature $j$ across the dataset. \\

\noindent
\textbf{Category Path Embedding:} 
We apply label encoding to assign a unique integer ID to each distinct path. A PyTorch nn. Embedding layer \cite{pytorch_embedding} maps each path ID to a dense vector of fixed dimension (e.g., 64).For each product, we compute the mean of its path embeddings to obtain a fixed-length vector.

\noindent
These embeddings are precomputed and fixed during model training. While not adaptive to tasks, this method offers a simple, scalable alternative to one-hot or multi-hot encoding, and extract semantic information from category hierarchies.\\

\noindent
\textbf{Final Node Features}

The final node embedded features $\mathbf{x}_i$ is formed by concatenating the individual feature vectors:
\[
\mathbf{x}_i = \left[ \mathbf{x}_i^{\text{title}} \, \| \, \mathbf{x}_i^{\text{group}} \, \| \, \mathbf{x}_i^{\text{numeric}} \, \| \, \mathbf{x}_i^{\text{cat}} \right],
\]

In our default configurations, $\mathbf{x}_i^{\text{title}}$ is a 384-dimensional SBERT embedding, and the full feature vector $\mathbf{x}_i$ is passed as input to graph neural network models without further transformation.

The final input feature vector for each node is constructed by concatenating the following components:

\begin{itemize}
    \item {SBERT-based title embeddings} (384 dimensions): captures the semantic meaning of product titles.
    \item {One-hot encoded group labels} (10 dimensions): represents the product domain (e.g., book, music, DVD).
    \item {Standardized numeric metadata} (7 dimensions): includes log-transformed sales rank, review counts, average ratings, and helpfulness scores.
    \item {Dense category path embeddings} (200 dimensions): encodes hierarchical product categories via an embedding layer.
\end{itemize}

This results in a final node feature vector of {601 dimensions} per product.

\vspace{0.5em}
\noindent\textbf{Optional Dimensionality Reduction}\\
To improve efficiency in low computational resource settings, we conducted optional dimensionality reduction of individual components. In particular, we reduced the SBERT embeddings dimension from 384 to 200 dimensions using Principal Component Analysis (PCA). This variant was not used in our primary experiments, but it is supported by our preprocessing pipeline and can be later integrated into future model iterations or different deployment scenarios.

\subsection*{2.5 Train/Test split}
To promote realistic generalization and avoid transductive leakage, we adopt an inductive train/test split strategy. Specifically, we randomly seperate the set of nodes into disjoint training and test subsets with 80:20 ratio. 

We then filtered the edge list to retain only those edges where both endpoints belong to the same split. This ensures that:
\begin{itemize}
    \item The model is trained only on a partial subgraph of the full graph.
    \item All evaluation is performed on entirely unseen nodes, simulating realistic cold start scenarios.
\end{itemize}

By verifying that there is no node overlap between the training and test sets, we were able to validate the split's inductiveness. This configuration reflects real-world limitations in recommendation systems that may eventually see the addition of new users or products.

\subsection*{2.6 Modeling:}
\subsection*{2.6.1 LightGCN Model}

LightGCN \cite{he2020lightgcn} is a simpler version of GCN that focuses on neighborhood aggregation by getting rid of feature transformation and nonlinear activation. We use the RecBole Library\cite{recbole} in our work.

\textbf{Input:}
The input is a user-item interaction matrix. Each entry indicates whether or not a user has interacted with an item.

\textbf{Leave-One-Out (LOO) Splitting:}  Each user's most recent interaction serves as the test (validation) case, while the remaining interactions are used for training. This ensures that the model's ability to predict how each user will interact with future items is tested.
Each user and item is assigned a unique integer ID during the processing.

\textbf{Layer 1 Embedding Layer:} 
Using embedding layers, the model maps user and item IDs to dense vectors of dimension embedding size (for instance, 64, 128, or 256). During training, these embeddings are learned and initialized at random.

\textbf{Layer 2 Graph Convolutional Layers:} 
Layer count: n\_layers (for instance, 1, 2, or 3 as a hyperparameter)
The embedding for each user and item is updated in each layer by aggregating (averaging) the embeddings of their neighbors in the user item interaction graph.
The final embedding for each user/item is the sum (or average) of the embeddings from all layers (including the initial embedding).

\textbf{Dropout:} Since the model is already simplified to prevent overfitting, the standard LightGCN architecture does not use explicit dropout.

\textbf{Output:} 
For a given user item pair, the model computes the dot product of their final embeddings to produce a score.
This score is passed through a sigmoid function to represent the probability of interaction (for binary classification tasks).
When rating or actually recommending, this can be used to sort by this probability, or manually use a threshold of 0.5 to convert to a 0/1 label.

\textbf{Training Configuration:}  
The model is trained using mini-batch stochastic gradient descent with the Adam optimizer.

\textbf{Optimization:} 
Adam optimizer with a learning rate chosen from {0.0005, 0.001, 0.005}.
Training batch size is selected from {512, 1024, 2048}.
Number of epochs: 50 or 100.

\textbf{Regularization:} 
L2 regularization (weight decay) is applied to the embeddings, with the regularization weight chosen from {0, 1e-5, 1e-4, 1e-3}.

\subsection*{2.6.2 GraphSAGE Model}
GraphSAGE (Graph Sample and Aggregate) is an inductive framework for learning node embeddings by aggregating information from the node's local neighborhoods\cite{hamilton2017graphsage}. It works well for recommendation and link prediction systems, particularly for dynamic and large-scale graphs. For real-world applications where new users and items are added on a regular basis, GraphSAGE's inductive ability to make predictions involving previously unseen nodes is essential. 

On a product co-purchase graph, we establish the task of product recommendation as an inductive link prediction task. Every node denotes a product and is linked to a dense feature vector $\mathbf{x}_i \in \mathbb{R}^d$, which is made up of pooled category path embeddings, standardized numeric metadata, one-hot group encoding, and SBERT title embeddings (refer to Section 2.1.3).

\textbf {Inputs} of the model are:
\begin{itemize}
    \item Node feature matrix: a matrix of node features.
    \item Graph Structure: the global edge index defines the structure of the graph used for neighborhood aggregation during GraphSAGE message passing.
    \item Edge pairs: node pairs (i, j) for which model predicts a link probability.
    \item Ground truth label: a subset of positive and negative labeled edges for training and evaluation.
\end{itemize}

\noindent
\textbf{Model Architecture}

Our model consists of two primary components: a GraphSAGE encoder and a multi-layer perceptron (MLP) link predictor (decoder) \cite{graphsage_simple_github, graphsage_python_tutorial}.

\begin{itemize}
    \item{GraphSAGE Encoder}  
The encoder is a two-layer GraphSAGE network, which aggregates features from a node’s local neighborhood to compute a latent node embedding. In the first layer, features from neighboring nodes are aggregated and passed through a ReLU activation, and the second layer transforms the outputs into the final embedding. This architecture supports inductive learning, allowing the model to generalize to previously unseen nodes.
    \item{Link Predictor/Decoder}  
To predict whether a link exists between a pair of nodes, we take the element wise product of their node embeddings and feed the result into an MLP. The MLP consists of multiple linear layers with ReLU activations and dropout, ending with a sigmoid output layer that produces a probability score between 0 and 1. Because this architecture supports inductive learning, the model can generalize to nodes that haven't been seen before\cite{deepgnn_link_pred}.
\end{itemize}
\noindent
\textbf{Mini-Batch Link Prediction with Neighbor Sampling}
\noindent
We utilize PyTorch Geometric's \texttt{LinkNeighborLoader} \cite{pyg_linkneighborloader} to scale the model to large graphs and lower GPU memory requirements. It performs efficient mini-batch training by:
\begin{itemize}
    \item Sampling a batch of training edges.
    \item Expanding a subgraph by sampling up to 10 neighbors at each of two hops for the edge endpoints.
    \item Computing the GraphSAGE embeddings only for this sampled subgraph.
\end{itemize}

This approach is capable to train on high resolution feature vectors and large graphs without loading the entire graph into memory at once. Here, we utilized a batch size of 1024 edges, which balances speed and memory efficiency. \\
\noindent
\textbf{Training Procedure}

The model is trained for up to 100 epochs using the Adam optimizer and L2 regularization with a tuned weight decay hyperparameter and dropout for regularization. We apply early stopping based on validation AUC with a patience of 20 epochs, meaning training stops if the validation performance does not improve for 20 consecutive epochs.

At each epoch, we evaluate the model on a validation set using the same mini-batching procedure. We report the best performing model based on validation AUC and record training time and classification metrics across epochs. After identifying the optimal hyperparameter configuration, we performed multiple runs with different random seeds to assess robustness and report averaged metrics with standard deviations.

\subsection*{2.6.3 Graph Attention Network (GAT)}
Graph Attention Networks (GAT) are a class of GNNs that leverage self attention mechanisms on graph structured data\cite{veličković2018graphattentionnetworks}. Unlike traditional graph convolutional networks, which assign equal or degree based weights to neighbors, GAT allows each node to learn unique attention weights for different neighbors. This attention mechanism addresses limitations of earlier spectral GNN approaches by avoiding costly matrix operations and enabling the model to operate in both transductive and inductive settings . GAT employs multi head attention, where multiple attention heads aggregate neighbor information in parallel and their outputs are concatenated (or averaged) to form the node’s new feature representation. This multi head design stabilizes learning and enriches the learned features, similar to the multi head attention in Transformers.

\vspace{0.5em}
\noindent
\textbf{Architecture:} \\
In our implementation, we use a two layer GAT model inspired by the original GAT paper \cite{veličković2018graphattentionnetworks}.
We utilize the PyTorch Geometric library \cite{fey2019fastgraphrepresentationlearning} for an efficient implementation of GATConv layers (which
leverages sparse GPU operations for speed). The model structure is as follows: \\
\textbf{Input Features:} 7-dimensional input feature vector per node (each item is described by 7 features).\\
\textbf{Layer 1:} Graph attention layer (GATConv) with hidden dimension 256 and 2 attention heads (concat=True). Each head produces 256 features; concatenation yields a 512-dimensional node embedding after the first layer. We apply ELU after this layer, and use dropout with $p=0.05$ both before the layer and inside the attention mechanism. The attention uses a LeakyReLU with negative slope $\alpha=0.2$ (PyG default).\\
\textbf{Layer 2:} Graph attention layer that maps the 512-dimensional input to a 512-dimensional output with heads=1 and concat=False (no concatenation), producing the final 512-dimensional node embedding. No additional activation is applied after the final layer.\\
\textbf{Dropout:} A modest dropout of 0.05 is applied to node features between GAT layers and within attention to regularize the model. Higher rates were observed to degrade performance, indicating limited overfitting. \\
\textbf{Output:} The model outputs 512-dimensional node embeddings. For binary link prediction, embeddings of two nodes $u$ and $v$ are combined via a dot product followed by a sigmoid to produce the edge probability. We initially experimented with a smaller output size (e.g., 16), but the implementation enforces $\textit{out\_dim}=\textit{hidden\_dim}\times\textit{heads}$ for the final layer; with our setting (256, 2 heads), this is 512, which we retained.\\
\vspace{0.5em}
\noindent
\textbf{Training Configuration:} \\
 We train for 150 epochs in an inductive setting. Negative sampling uses a negative:positive ratio of 0.3, and the positive class is weighted by 2.5. We use Focal Loss (use\_focal=True) with $\gamma=2.0$; early stopping patience is 20 epochs.\\
\vspace{0.5em}
\noindent
\textbf{Optimization:} \\
Adam optimizer with initial learning rate 0.005 and weight decay $1\times 10^{-4}$. 
A MultiStepLR scheduler decays the learning rate at epochs 60, 100, and 140 with decay factor $\gamma = 0.1$.

\subsection*{2.6.4 PinSAGE Model}

PinSAGE \cite{ying2018pinsage} is a GNN architecture designed to scale recommendation tasks to web scale item graphs. Inspired by GraphSAGE, PinSAGE combines random walk based neighbor sampling and learned embeddings to generate expressive node representations. Implementation for this project adapts PinSAGE to an inductive link prediction setting with precomputed random walks to accelerate neighbor sampling during training . 

Following stages are implemented in modeling pipeline:
\begin{itemize}
  \item \textbf{Precomputation:} For each node, run 15 random walks of length 5 to generate a personalized neighborhood with importance based sampling probabilities. These precomputed neighbors and probabilities were saved as dictionaries for both training and validation graphs.
  
  \item \textbf{Sampler:} At training time, loaded the precomputed walk dictionaries and implemented a customized \texttt{PrecomputedSampler} which performed layered sampling using importance weights. For each mini-batch, blocks were constructed by recursively sampling up to 3 hop neighbors, forming the computation graph.
  
  \item \textbf{Model Input:} The model received pairs of item nodes (positive and negative samples), sampled neighbors from the graph, and passed the resulting blocks into a GNN encoder followed by a dot product decoder.
  
  \item \textbf{Training Objective:} The model was trained to distinguish between co-purchased and non co-purchased item pairs with binary cross entropy loss function.
\end{itemize}

\vspace{0.5em}
\noindent
\textbf{Model Architecture:} 

Implemented a three layer PinSAGE style GNN with customized  GraphSAGE layers and dot product decoder:

\begin{itemize}
  \item \textbf{Input:} 1937-dimensional node features (four 384-dimensional SentenceTransformer encoded text embeddings and seven numerical features).
  \item \textbf{Layer 1:} DenseGraphSAGE with hidden size 256 and ReLU activation.
  \item \textbf{Layer 2:} DenseGraphSAGE with output size 128 and dropout (0.25).
  \item \textbf{Layer 3:} DenseGraphSAGE with output size 128 and dropout (0.25).
  \item \textbf{Decoder:} DotProductPredictor computes edge scores as $\text{dot}(h_u, h_v)$ between node pairs.
\end{itemize}

\textbf{Unique Design Points:}
\begin{itemize}
  \item Importance based neighbor sampling using precomputed random walks.
  \item Customized 3 hops sampling through a dynamic \texttt{PrecomputedSampler}.
  \item Fully inductive train/test split, with unseen nodes tested during validation.
\end{itemize}

\noindent
\textbf{Training Setup:} \\
The model was trained for 30 epochs with the Adam optimizer:
\begin{itemize}
  \item \textbf{Learning rate:} 1e-4, \quad \textbf{Weight decay:} 4e-5
  \item \textbf{Batch size:} 512, \quad \textbf{Dropout:} 0.25
  \item \textbf{Loss function:} \texttt{BCEWithLogitsLoss} with balanced positive/negative sampling
\end{itemize}
Model performance was evaluated using AUC and Average Precision (AP). The best validation AUC was used to select the final best model.

\noindent
\textbf{Output:} \\
The model output learned node embeddings for items. For each (head, tail) node pair, a dot product score was computed to predict co-purchase likelihood. Final model weights and training plots were saved for downstream evaluation and analysis.

\section{Experiments and Results}

\subsection*{3.1 Loss Function:} Binary Cross Entropy with Logits (BCEWithLogitsLoss) was used, with a positive class
weight of 2.0 to address class imbalance. This effectively penalizes missing a true edge twice as
much as a false alarm, focusing the model on the sparser positive class. We also experimented with
the Focal Loss [3] ($\gamma=2$) as an alternative imbalance handling technique, which down weights easy negatives and focuses on hard examples \cite{lin2018focallossdenseobject} . However, we found that a properly set pos\_weight in BCE was sufficient in our case, so we did not ultimately deploy focal loss in the final runs.
\\\\
\textbf{Early Stopping: }We monitored validation AUC for early stopping with a patience of 20 epochs. In
practice, the validation AUC kept improving until very late in training, so the model trained for the
full 150 epochs (no early stop triggered).\\\\
\textbf{Negative Sampling: }For training the link prediction task, we used a negative sampling ratio of 0.3 –
meaning we sampled fewer non edge pairs relative to true edges each epoch to maintain a
reasonable class balance in each batch. This, combined with the pos\_weight, helped the model not
to be overwhelmed by the many possible negative edges.

\subsection*{3.2 Success Metrics}
\begin{itemize}
  \item \textbf{AUC} measures the model's ability to distinguish between positive and negative item pairs. It is robust to class imbalance and commonly used in binary classification settings.
  \item \textbf{AP} summarizes the precision recall curve and reflects the model’s ability to rank positive pairs higher. It is especially valuable in real world use cases where top-$K$ item recommendations are required—such as displaying related products on an Amazon product page. 
  \item \textbf{Loss} refers to the binary cross entropy loss computed between predicted and true labels. It provides a direct optimization target during training.
\end{itemize}

\subsection*{3.3 Hyperparameter Searching Strategy}
\textit{Random search} is applied to explore the hyperparameter space in this project. Randomly selected parameter sets from predetermined ranges were tested, rather than thoroughly analyzing every possible configuration. We assessed the model's AUC and AP following each trial, using the findings to direct the subsequent sampling area. We were able to concentrate on promising regions of the parameter space thanks to this adaptive approach. Random search offered a balance between efficiency and exploration, especially considering the high computational cost of training of GNN models.

\subsection*{3.4 Model Performance}
\begin{table*}[ht]
\centering
\begin{tabular}{l c ccc ccc c}
\toprule
\textbf{Model} & \textbf{GPU} & \multicolumn{3}{c}{\textbf{Training}} & \multicolumn{3}{c}{\textbf{Validation}} & \textbf{Epoch Time (s)} \\
 & & AUC & AP & Loss & AUC & AP & Loss & \\
\midrule
LightGCN  & Nvidia H200 – 141G         & 0.9094 & 0.9197 & 0.0025 & 0.8355 & 0.8592 & 0.9591 & 569.17 \\
GraphSAGE & Nvidia L4 – 24G     & 0.9981 & 0.9979  & 0.0506 & 0.9976 & 0.9974 & 0.0594 & 19.60 \\
GAT & Nvidia L4 – 24G & 0.9719 & 0.9945 & 0.0843 & 0.9729 & 0.9888 & 0.1283 & 2.29 \\
PinSAGE   & Nvidia A100 – 40G   & 0.9298 & 0.9434 & 0.4493 & 0.9562 & 0.9649 & 0.4306 & 1881.00 \\
\bottomrule
\end{tabular}
\caption{Performance and runtime comparison of GNN models on the Amazon co-purchase dataset. Each model was evaluated using both training and validation metrics, along with GPU type and per-epoch training time.}
\label{tab:model-comparison-gpu}
\end{table*}

\subsection*{3.4.1 LightGCN}

\paragraph{Experimental Setup}

We trained a LightGCN model using the RecBole framework on our Amazon product co-purchasing dataset. The experimental setup is designed to evaluate the effectiveness of LightGCN for product recommendation tasks.

\paragraph{Best Configuration and Hyperparameter Tuning}

We performed extensive hyperparameter tuning using random search rather than grid search, which allowed us to explore a diverse set of configurations efficiently. This approach provided broad coverage without excessive computational cost.

The hyperparameter search space included:
\begin{itemize}
    \item Embedding dimensions: \{64, 128, 256\}
    \item Graph convolution layers: \{1, 2, 3\}
    \item Learning rates: \{0.01, 0.005, 0.001\}
    \item Batch sizes: \{512, 1024, 2048\}
    \item Reg weight: \{0, 1e-5, 1e-4, 1e-3\}
\end{itemize}

Each sampled configuration was trained with early stopping based on validation AP. The best configuration, selected after evaluating many trials, was:

\begin{itemize}
    \item Embedding dimensions: \{128\}
    \item Graph convolution layers: \{2\}
    \item Learning rates: \{0.001\}
    \item Batch sizes: \{1024\}
    \item Reg weight: \{1e-3\}
\end{itemize}

This configuration offered excellent performance while fitting within the memory limits of a NVIDIA H200 GPU.

Our best model converged quickly, with early stopping triggered at epoch 30. The validation metrics at this point were:

\begin{itemize}
    \item \textbf{Recall@10:} 0.5731
    \item \textbf{NDCG@10:} 0.4528
    \item \textbf{AP:} 0.8592
    \item \textbf{AUC:} 0.8355
\end{itemize}

\begin{figure}
    \centering
    \includegraphics[width=1\linewidth]{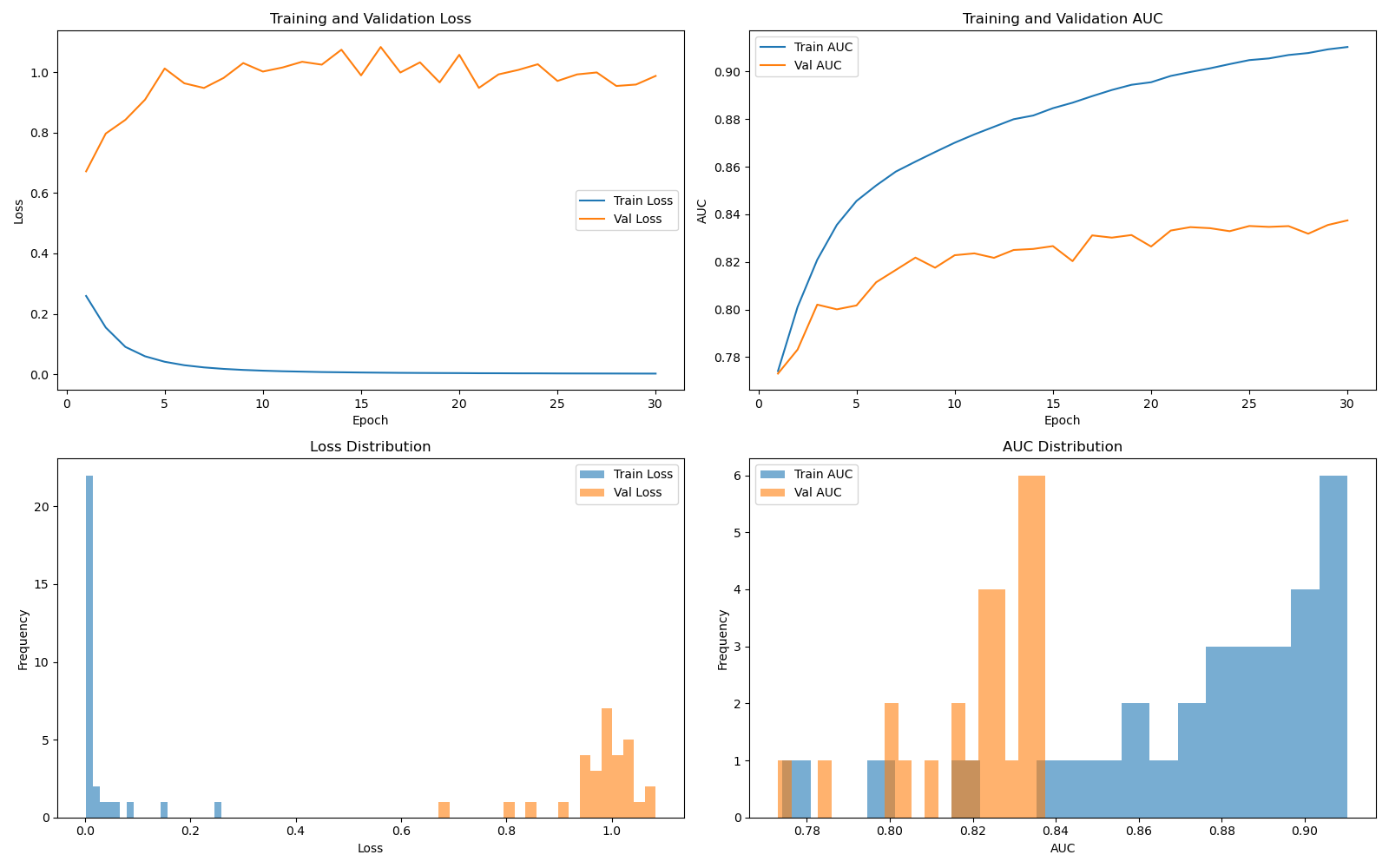}
     \caption{
Training and validation dynamics of LightGCN over a single representative run, 30 epoch.
(Top-left) Training and validation loss curves, showing the model’s convergence behavior.
(Top-right) Training and validation AUC curves, indicating the evolution of discriminative performance.
(Bottom-left) Histogram of loss values across epochs, illustrating the distribution of optimization errors.
(Bottom-right) Histogram of AUC values across epochs, reflecting consistency in predictive quality.
}
  \label{fig:lightGCN-training}
\end{figure}

These results indicate generalization performance in an inductive setting.  The training process was shown in Figure ~\ref{fig:lightGCN-training}. For more training metrics, please refer to Table~\ref{tab:model-comparison-gpu}.

However, unlike traditional deep learning models, LightGCN does not employ nonlinear activation functions or feature transformation layers, and it propagates embeddings across multiple layers purely through neighborhood aggregation. While AUC provides a global view of classification performance, it fails to capture the ranking nature and extreme imbalance of recommendation systems. Metrics that emphasize ranking quality and top-K performance are more meaningful for evaluating recommendation effectiveness.

In practice, we observe that the validation loss (e.g., BPR loss) may not correlate well with ranking performance metrics such as Recall or NDCG. This is because the optimization objective (pairwise loss) does not directly reflect top-k ranking quality, and the simplistic architecture of LightGCN leads to overfitting the training pairs without improving generalization.

\subsubsection*{3.4.2 GraphSAGE}
\noindent
\textbf{Experimental Setup}

We trained a 2-layer GraphSAGE encoder combined with a multi-layer perceptron (MLP) link predictor using PyTorch Geometric’s \texttt{LinkNeighborLoader}. An inductive train/test node split ensured that test nodes were completely unseen during training, emulating cold-start conditions in real-world recommendation systems.

Training was conducted with mini-batch neighborhood sampling (2-hop, [15, 10]), a batch size of 1024, and binary cross-entropy loss with a computed positive class weight. Node features consisted of SBERT encoded titles, one-hot group labels, standardized numeric metadata, and pooled category path embeddings.\\

\noindent
\textbf{Best Configuration and Hyperparameter Tuning}\\
We performed hyperparameter tuning using random search instead of grid search so that we could test a wide set of configurations with limited computational resources efficiently. This approach provided a broad coverage without excessive computational cost.

The hyperparameter search space included:
\begin{itemize}
    \item Hidden dimensions: \{32, 64, 128\}
    \item Number of MLP layers: \{2, 3\}
    \item Dropout rates: \{0.3, 0.5\}
    \item Learning rates: \{0.01, 0.005, 0.001\}
    \item Batch sizes: \{512, 1024\}
    \item Neighbor sampling fan-out: \{[5, 5], [10, 10], [15, 10]\}
    \item weight decay: \{0.0, 1e-6, 1e-5, 5e-5, 1e-4\}
\end{itemize}

Each sampled configuration was trained with early stopping based on validation AUC. The best configuration was:

\begin{itemize}
    \item {Hidden dimension:} 32
    \item {Number of MLP layers:} 2
    \item {Dropout rate:} 0.3
    \item{Learning rate:} 0.001
    \item {Batch size:} 512
    \item{Neighbor sampling:} [5, 5]
    \item {Weight decay:} 1e-4
\end{itemize}

This suggests that a relatively simple model architecture (32 hidden dimension with two MLP layers) with fewer parameters helped reduce overfitting while still capturing relevant structural information from the Amazon co-purchase graph. The dropout rate of 0.3 provides moderate regularization without excessively discarding information. A low learning rate (0.001) stabilizes optimization, allowing the model to converge smoothly, while a weight decay of 1e-4 further constrains parameter growth and improves generalization. A moderate batch size (512) is efficient and stable gradient updates during mini-batch training. Finally, sampling a narrow neighborhood [5,5] enabled the model to access enough local context without excessive computational overhead.Together, these choices yield a model that generalizes well to unseen nodes in the inductive setting while maintaining training efficiency.

This configuration offered excellent performance while with the memory limits of the NVIDIA L4 GPU. Smaller hidden dimension yielded strong performance, probably due to the semantic richness of the SBERT-based input features.

Our best model converged quickly, with early stopping triggered at 30-50 epoch in our replicate experiments (n=5). The validation metrics at this point were:

\begin{itemize}
    \item \textbf{Train Loss:} $0.0506 \pm 0.00095$
    \item \textbf{Train Acc:} $0.9826 \pm 0.00040$
    \item \textbf{Train AUC:} $0.9981 \pm 0.00009$
    \item \textbf{Train AP:} $0.9979 \pm 0.00009$
    \item \textbf{Val Loss:} $0.0594 \pm 0.00150$
    \item \textbf{Val Acc:} $0.9792 \pm 0.00040$
    \item \textbf{Val AUC:} $0.9976 \pm 0.00004$
    \item \textbf{Val AP:} $0.9974 \pm 0.00004$
    \item \textbf{Time (s):} $19.60 \pm 0.673$
\end{itemize}

\begin{figure}
    \centering
    \includegraphics[width=1\linewidth]{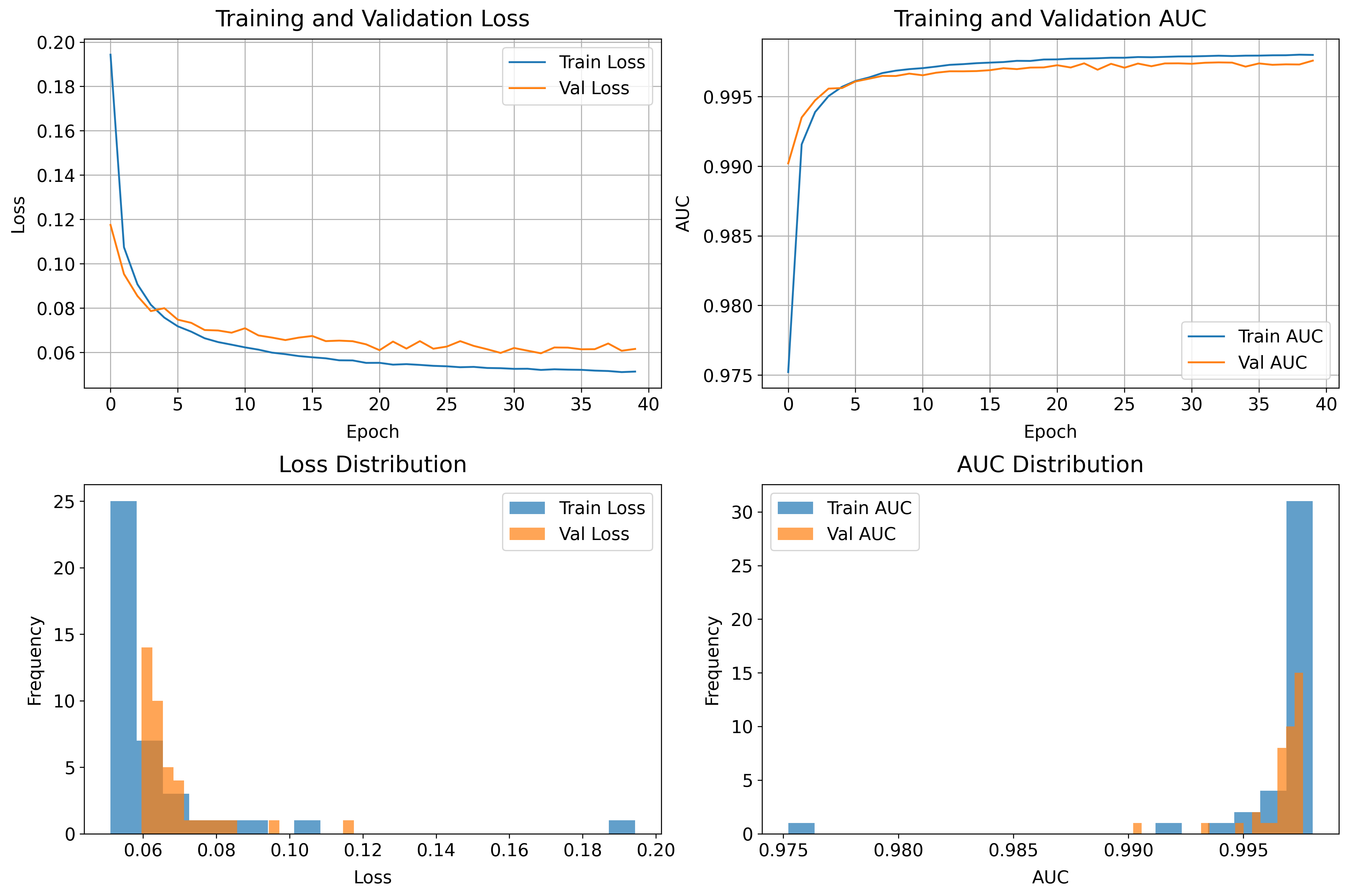}
    \caption{Training and validation dynamics of GraphSAGE. 
Plots are arranged as in Figure.~\ref{fig:lightGCN-training}, showing faster convergence,lower final loss, and higher AUC/AP compared to LightGCN, with minimal train–validation loss gap (0.009).}
    \label{fig:graphsage-training}
\end{figure}

One representative training process is shown in Figure~\ref{fig:graphsage-training}. 
These results indicate a good generalization performance in the inductive setting. Performance was stable across epochs, and both training and validation Loss (upper left) and AUC curves (upper right) showed smooth convergence after 30~50 epochs without overfitting.

Our model satisfies our objectives for scalability, generalization, and practicality. We created a high-performing GNN pipeline for inductive link prediction by utilizing neighborhood sampling, pretrained features, and random search tuning.

GraphSAGE demonstrated its effectiveness and suitability for resource-constrained environments by achieving the highest AUC of all the modes while maintaining a low training time (16.19s per epoch) on a comparatively lightweight L4 GPU (Table~\ref{tab:model-comparison-gpu}).

In addition, we used the A100 GPU for training, which should provide more computational capacity than the L4. But when we used it for GraphSAGE (data not shown here), training speedup is minimal. This implies that the expense of repeatedly sampling subgraphs and moving them from the CPU to the GPU, rather than GPU compute, limits our training pipeline. This bottleneck has been widely recognized in prior system level analysis \cite{ma2019neugraph, zhu2019aligraph}. Since this happens during each mini-batch, it becomes the main bottleneck, limiting the benefits of using a more powerful GPU like the A100.

\subsection*{3.4.3 GAT}
The experiments were conducted on a large graph dataset derived from an e-commerce application. The
graph consists of approximately 402,691 nodes (representing items/products) and 808,185 edges
representing similarity relationships. We performed a inductive split of edges into training/validation/test sets (ratio roughly 70\%/15\%/15\%). Specifically, about 790,854 edges were used for
training, 36,294 edges for validation, and 36,016 edges for testing (each true edge has an equal number of
sampled negative edges in these sets) . In a inductive setting, all nodes are present during training;
only edges are withheld for validation/test, meaning the model can learn embeddings for all nodes. We
opted for this approach to fully utilize item information in training.

Training was executed on a single NVIDIA L4 GPU with 24 GB memory. Thanks to the efficient PyG library
and our mini-batch negative sampling strategy, each epoch was fast – about 2.3 seconds per epoch on
average – and the entire 150 epochs completed in ~7.1 minutes of wall clock time. Sparse GAT operations occupied only about 22 GB of the available 24 GB of GPU memory, demonstrating their efficient memory utilization. This speed and scalability underscore the advantage of using PyTorch Geometric’s optimized kernels for graph data\cite{fey2019fastgraphrepresentationlearning}.\\
\begin{figure}
    \centering
    \includegraphics[width=1\linewidth]{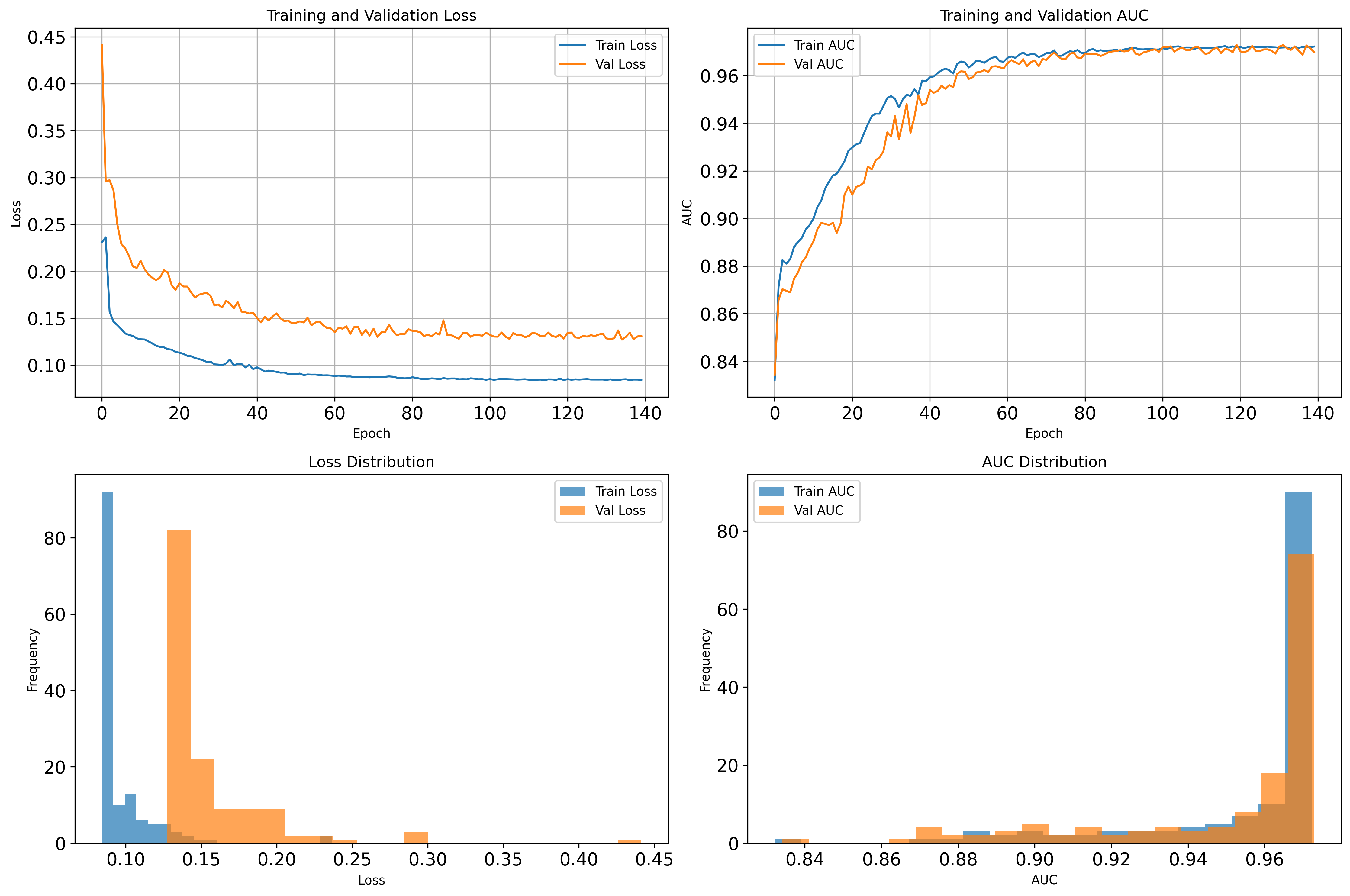}
    \caption{Training and validation dynamics of GAT. 
Using the same plot layout as Figure.~\ref{fig:lightGCN-training}, the model converges more slowly over 150 epochs, achieves strong final AUC/AP, but exhibits mild overfitting, with a moderate train–validation loss gap of 0.045.}
    \label{fig:gat_training}
\end{figure}
Training history of the GAT model is shown in Figure~\ref{fig:gat_training}, present training vs. validation loss (left) and AUC (right) over
150 epochs. Both training and validation metrics improve steadily. The loss curves (left) drop sharply in the
first ~10–20 epochs and flatten out by epoch ~50, with the validation loss closely tracking the training loss
throughout. The AUC curves (right) start around 0.84 and climb rapidly; the validation AUC reaches ~0.94 by
epoch 30 and then improves more gradually. By epoch ~100, the model achieves over 0.97 AUC on both
training and validation, and it inches up to ~0.9729 by epoch 120. Notably, the training and validation curves
are very close together for both loss and AUC, indicating minimal overfitting – the GAT generalized almost
as well as it learned on the training data.

At the end of training, we evaluate the model on the test set (held out edges). The GAT achieved a final test
AUC of 0.9723 and Average Precision (AP) of 0.9889, with a test loss of ~0.1287. An AUC of 0.9723 indicates that 97.23 percent of the time, the model can correctly distinguish a true edge vs. a non-edge, a very
high level of performance for this task. The AP of 0.9889 is likewise extremely high, reflecting that the
precision recall curve is near the top in most regions (the model misses very few true edges and has a low
false positive rate). During training, the best validation AUC reached ~0.9729 (around epoch 120 ), which is essentially on par with the test AUC, confirming that the model’s performance generalizes well to unseen
edges. 

In terms of efficiency, the full-batch GAT training on this graph was feasible and fast. Each epoch processed
over 790,854 training edges (plus a 30\% of negative samples) in ~2.3 seconds. The cosine
learning rate schedule helped stabilize training in the later epochs-as the learning rate became small, the
training and validation losses flattened out and AUCs oscillated only slightly (see Figure~\ref{fig:gat_training}, after epoch 100).
This scheduling avoided any sudden spikes or divergence that might occur with a fixed or step-based
learning rate, at the cost of a longer training time to converge the last few thousandths in AUC. Given the
already short training time, this trade-off was acceptable. We also note that the lightweight architecture
(two layers, relatively low dropout) and strong regularization (weight decay, sampling) contributed to the
model’s ability to fit the data well without overfitting. The final model’s attention weights can be interpreted
to some extent, e.g., we could inspect which neighbor nodes each item attended to most, potentially yielding insights, though such analysis is beyond the scope of this report.\

\subsection*{3.4.4 PinSAGE Model}

\paragraph{Training Setup \& Hardware Constraints:}We trained PinSAGE \cite{ying2018pinsage} model on Google Colab with NVIDIA A100 GPU. Because of the complex structure of the model, its dependence on precomputed random walks, and the expense of importance sampling, each training epoch took roughly \textbf{30 minutes}, with a total runtime of more than \textbf{15 hours} for 30 epochs. Our ability to perform complete hyperparameter tuning was severely limited by these runtime and resource constraints.
\paragraph{Best Model Configuration.}
Random search over a constrained configuration space is applied to identify a high performing setup. The best performing configuration was:
\begin{itemize}
  \item \texttt{num\_walks = 15}, \texttt{walk\_length = 5}, \texttt{num\_neighbors = 8}, \texttt{num\_layers = 3}, \texttt{batch\_size = 512}, \texttt{num\_epochs = 30}, \texttt{learning\_rate = 0.0001}, \texttt{hidden\_feats = 256}, \texttt{out\_feats = 128}, \texttt{dropout = 0.25}, \texttt{train\_ratio = 0.8}
\end{itemize}
\paragraph{Hyperparameter Insights:}
\begin{itemize}
  \item \textbf{Random Walks \& Sampling:} Richer context is captured by more walks and neighbors, but memory and training time are significantly boosted. To strike a balance between expressiveness and efficiency, we used 15 walks of length 5 with 8 sampled neighbors.
  \item \textbf{Hidden Dimensions \& Layers:} The representational power was enhanced by larger hidden dimensions (256). And effective multi-hop aggregation  was achieved by three GNN layers.
  \item \textbf{Dropout and Learning Rate:} A dropout rate of 0.25 and a low learning rate (1e-4) contributed to steady convergence while avoiding overfitting.
  \item \textbf{Batch Size:} A size of 512 enabled stable training while fitting within memory limits of the A100 GPU.
\end{itemize}
\paragraph{Quantitative Results:}
The best model was selected based on validation AUC, peaking at \textbf{Epoch 25}. Final results:

\begin{itemize}
  \item \textbf{Training AUC:} 0.9339 \quad \textbf{Training AP:} 0.9473
  \item \textbf{Validation AUC:} \textbf{0.9562} \quad \textbf{Validation AP:} \textbf{0.9649}
  \item \textbf{Training Loss:} 0.4412 \quad \textbf{Validation Loss:} 0.4306
\end{itemize}

Validation metrics steadily improved over all 30 epochs. Since no performance plateau was observed, further improvements may have been achievable with extended training. Unfortunately, we were constrained by runtime and compute limitations.

\paragraph{Qualitative Analysis from Training Curve:}
\begin{itemize}
  \item \textbf{Loss Trends:} Training and validation loss both steadily decreased with minimal gap which indicated effective generalization and limited overfitting.
  \item \textbf{AUC \& AP Trends:} AUC and AP improved consistently without oscillation, reaching high values around epoch 25--30. This confirms the model’s ability to learn discriminative embeddings and rank edges effectively.
  \item \textbf{Room for Improvement:} No clear saturation was observed in validation AUC or AP, suggesting that the model had not yet fully converged. More epochs could further improve performance without computation resource limitation.
\end{itemize}

\begin{figure}[h]
    \centering
    \includegraphics[width=1\linewidth]{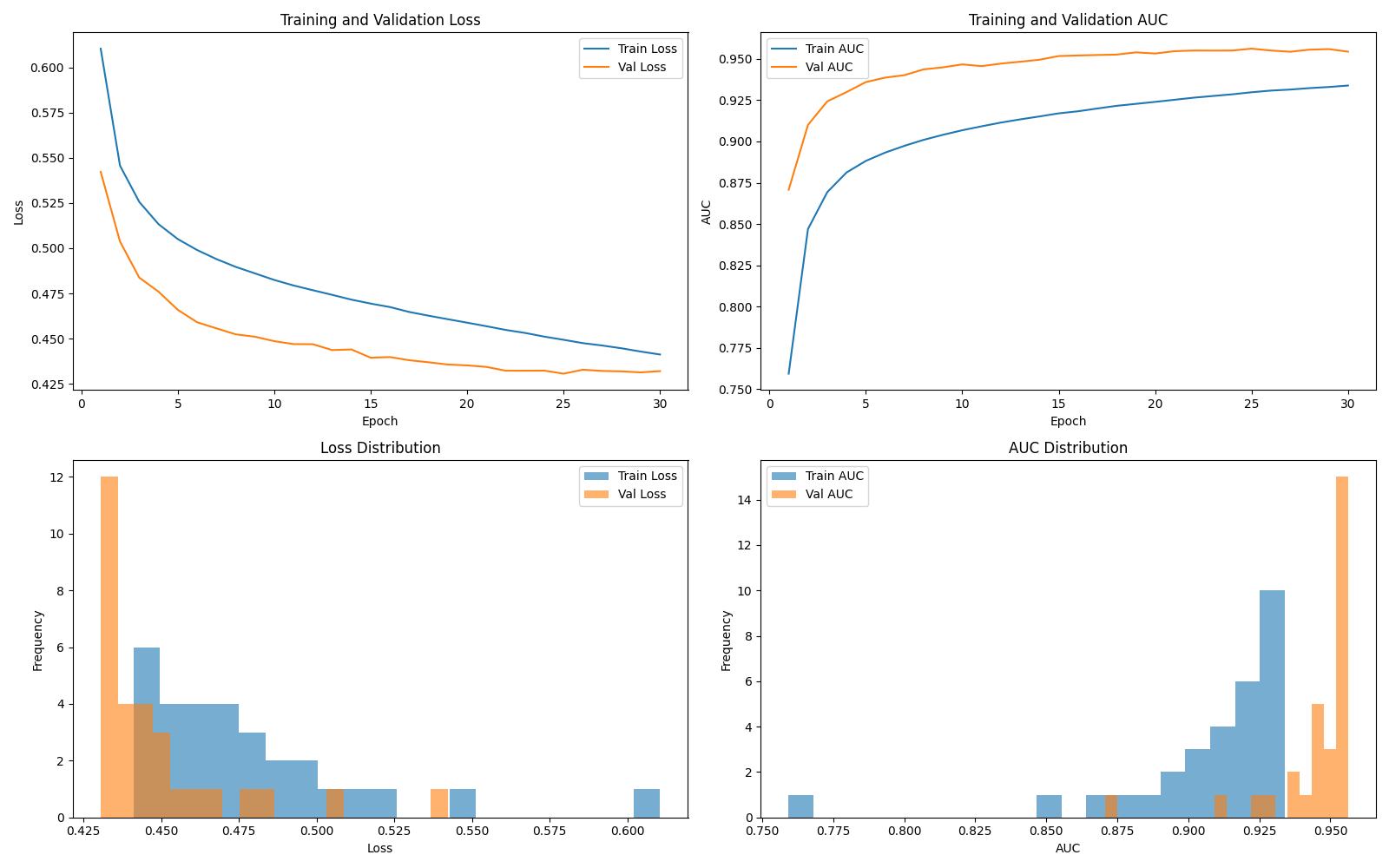}
    \caption{Training and validation dynamics of PinSAGE. 
Following the format of Fig.~\ref{fig:lightGCN-training}, the model converges within 30 epochs,
but with slightly higher final loss and moderate AUC/AP
Training is notably slower due to complex neighbor sampling.}
    \label{fig:pinsage-training}
\end{figure}

\paragraph{Model Comparison \& Challenges:}
Although the PinSAGE model achieved strong performance, it did not outperform our GAT and GraphSAGE baselines. Following hypothesis may contribute to this scenario:

\begin{itemize}
  \item \textbf{Tuning Difficulty:} PinSAGE is more sensitive to hyperparameters due to its complex sampling pipeline and walk based features. Exhaustive tuning was infeasible under computational resource constraints.
  \item \textbf{Long Runtime:} Each training run spanned 15 hours, making repeated trials impractical.
  \item \textbf{Feature Redundancy:} Random walks may introduce noisy or redundant neighbor context, especially in sparse or noisy graphs, which can hurt generalization.
\end{itemize}

\begin{figure}
      \centering
      \includegraphics[width=1\linewidth]{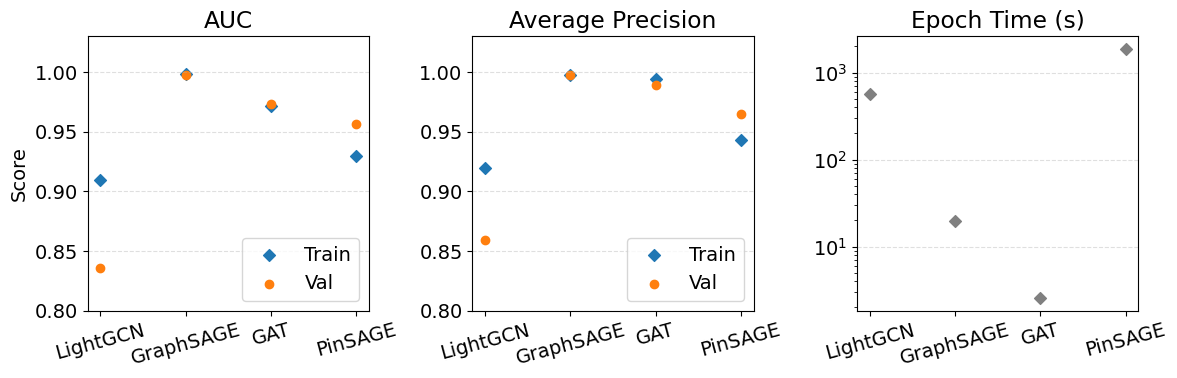}
      \caption{Performance comparison of GNN models on AUC, Average Precision, and per-epoch training time. \\ *GPU for different model varies, please refer to Table 4.  }
      \label{fig:placeholder}
  \end{figure}

\section{Discussion}
\paragraph{Transductive vs Inductive Evaluation:} In early experiments, we observed that using a transductive or edge based train/test split (where nodes appear in both sets) resulted in unusually high validation accuracy and AUC before any training had occurred. This led us to suspect the data leakage, where the model effectively memorized node identities or neighborhood overlaps.

This issue occurs because, in transductive settings, the node features and embeddings are shared between training and test phases, allowing the model to indirectly benefit from information about test nodes during training. Similar concerns have been identified in prior studies~\cite{fey2020gnnbenchmark,salha2021evaluating}, which highlight how transductive settings allow node features and embeddings to be shared across training and test phases, thus enabling indirect access to test node information during training. To mitigate this, we adopted a strict inductive splitting strategy in which the node set was partitioned prior to edge filtering, ensuring that no test node was present during training. This change significantly improved the robustness of our evaluation and produced learning dynamics more consistent with real world deployment scenarios.

To address this problem, we adopted a strict inductive split where nodes are partitioned before edge filtering, ensuring that no test node is seen during training. This dramatically improved the realism and robustness of our evaluation, and revealed more meaningful real-life learning dynamics.

\paragraph{Negative Sampling Ratio:} An initial configuration with a negative sampling ratio of 10:1 to provide the model with a strong signal from negative examples. However, this configuration quickly consumed GPU memory and could not run even a single epoch. Reducing the ratio to 3:1 made training feasible, but led to significantly degraded validation and test performance. After multiple rounds of tuning, lower negative sampling ratios (e.g., 1:1) resulted in better generalization. This suggests that aggressive negative sampling may introduce noisy or redundant gradients that harm learning, whereas a more balanced ratio leads to cleaner optimization and improved test time performance.

\paragraph{Scalability and Resource Efficiency:} Training GNNs on large scale graphs with high-dimensional node features presents both algorithmic and hardware challenges. GraphSAGE, while originally more general purpose, faces similar constraints when applied to large graphs. In our experiments on the Amazon co-purchase network, training GraphSAGE on the full graph required frontier GPUs (NVIDIA A100) due to memory demands from SBERT-based node features and dense edge connectivity. This level of hardware is not always accessible in academic or production environments.
  
To make training feasible and reproducible on widely available GPUs such as NVIDIA T4 or L4, we adopted a mini-batch training strategy similar to PinSAGE. Specifically, we used PyTorch Geometric's \texttt{LinkNeighborLoader} to sample 2-hop subgraphs during training, allowing the model to scale while maintaining strong performance. This design allowed us to train effectively without requiring full graph access or excessive compute resources. This design allows separation of model performance from specialized hardware availability, making the experiments more accessible to a broader research community.

\paragraph{Installation:} Installing CUDA-compatible versions of PyTorch, DGL and other necessary libraries is necessary to use GPU in training and validation steps. However, it was challenging to resolve dependency conflicts among various versions of PyTorch, DGL and CUDA because many combinations resulted in runtime incompatibilities and installation errors. Early experimentation was severely affected by these problems, which had to be fixed by careful version pinning and manual installation in a controlled setting.  

\paragraph{Model Comparison:}
We evaluated three expressive GNN models— LightGCN, GraphSAGE,  GAT and PinSAGE on the Amazon co-purchase network using an inductive link prediction setup. All three models successfully leveraged neighborhood structure and node features, but differed in generalization performance, computational efficiency, and hardware requirements. The results are display in Table 2 and Figure 5.

GraphSAGE achieved the highest overall validation performance, with an AUC of 0.9977 and AP of 0.9974, representing a substantial improvement over the initial project results. Refinements to the model architecture and training pipeline contributed to both higher accuracy and stable convergence, though at a moderate computational cost (19.54 s per epoch on an Nvidia L4 GPU). Further efficiency gains could be achieved through graph coarsening ~\cite{convmatch2023} or asymmetric modeling approaches~\cite{aml2023} that reduce redundant computations during message passing. These results highlight GraphSAGE’s strength for inductive link prediction when appropriately tuned.

GAT remained the fastest model evaluated, converging in only 2.29 s per epoch on an Nvidia L4 GPU, while still delivering strong validation performance (AUC 0.9729, AP 0.9888). This efficiency stems from data-loading optimizations, effective batching, and refined regularization strategies. The accuracy gap relative to GraphSAGE indicates potential for improvement, for instance via hybrid architectures that deepen neighborhood aggregation or by enriching node features if additional modalities become available in future datasets. Although GAT no longer achieved the best accuracy, its combination of speed and solid predictive quality makes it attractive for latency-sensitive applications. 

PinSAGE, while architecturally sophisticated and designed for web-scale recommendation systems, showed middling performance in our task. It achieved a validation AUC of 0.9562 and AP of 0.9649, but suffered from extremely long training times (over 1800 seconds per epoch). This inefficiency may be due to the complexity of its personalized neighbor sampling and dense feature transformations. While effective, PinSAGE's computational cost limits its practical use without aggressive optimization or large-scale infrastructure. This bottleneck may be alleviated by optimized sampling strategies or communication-efficient distributed training~\cite{splpg2025}.

While LightGCN is a popular model for collaborative filtering, it underperformed on our link prediction task. Despite achieving low training loss, its validation AUC and AP were considerably worse than all other models (Table 2). We attribute this to LightGCN’s design, which omits feature transformation and non-linearity, making it less suited for tasks involving rich node features such as SBERT embeddings. Furthermore, LightGCN does not leverage edge weights or attention mechanisms, which may have limited its expressiveness on the co-purchase network, where fine-grained item similarity patterns matter. These results suggest that LightGCN's assumptions is primarily designed for sparse, implicit interaction matrices—do not align well with our problem setting.

Overall, our experiments revealed clear trade-offs among the evaluated GNN models. GraphSAGE achieved highest predictive performance but a moderate computational cost, suggesting the efficiency optimization could further enhance its feature. GAT provided an outstanding speed with strong accuracy, but left potential for accuracy oriented refinements. PinSAGE demonstrated competitive accuracy but required prohibitively long training times, indicating that more efficient neighbor sampling strategies could improve its viability. LightGCN’s underperformance in this feature-rich setting points to potential gains from incorporating richer feature transformations or hybrid architectures.

 While our evaluation focused on representative architectures, existing research offers complementary strategies—such as coarsening, asymmetric modeling, multi-modal fusion, and temporal adaptation—that could address many of the limitations observed here. We outline these directions in the following section on Future Work.

\section{Conclusions and Future Work}
In this project, we investigated graph neural network architectures for inductive link prediction on the Amazon co-purchase network. We focused on models that could scale to large graphs with rich node features, including LightGCN, GraphSAGE, GAT and PinSAGE.  

Our current category path representation preserves ordering via label encoding and embedding layers but does not explicitly leverage the category graph structure. Preliminary experiments with graph-based embedding methods  showed potential for capturing richer inter-category relationships, though tuning was limited by time constraints. Future work could refine this approach by optimizing random-walk parameters, incorporating co-occurrence edges, and applying regularization to produce more semantically coherent category clusters.

Looking ahead, future work could explore:  
1) \textbf{Interpretability:} Analyzing learned attention weights in GAT or feature importances in GraphSAGE to better understand model decision-making.  
2) \textbf{Feature enrichment:} Incorporating hierarchy-aware or graph-based category embeddings, and, when available, integrating true multi-modal inputs such as product image embeddings to provide complementary signals for cold-start scenarios.  
3) \textbf{Efficiency improvements:} Profiling GPU utilization and experimenting with alternative neighbor sampling methods, graph coarsening, or mixed-precision training to reduce runtime and memory costs.  
4) \textbf{Cold-start and dynamic settings:} While our inductive node split already evaluates static cold-start performance on unseen products, future extensions could address \emph{continuous} cold-start scenarios where new items are introduced over time, as well as \emph{dynamic graph} settings where co-purchase patterns evolve. Such extensions may require temporal GNN architectures or incremental embedding updates.

Overall, our findings highlight the need to balance model complexity, predictive performance, and computational cost when deploying GNNs in real-world recommendation environments, and they identify concrete pathways for improving both feature representations and system efficiency.

\section{Author Contributions}
M. Cao and F. F. Yang jointly led the data preparation. M. Cao coordinated project discussions and integration efforts, and implemented the PinSAGE pipeline, including graph construction and sampling. F. F. Yang developed reusable graph code, implemented the GAT model, and refined it in the later stage. Y. Jin developed the feature embedding pipeline and GraphSAGE workflow, refined the GraphSAGE model, and led later-stage manuscript revisions, adding figures, tables, and expanded sections. Y. Yan prepared the review dataset and implemented the LightGCN pipeline. All authors contributed to project design, experimentation, analysis, , manuscript preparation and Github repository.

\section{Acknowledgment}
This work originated as a final project for CS7643: Deep Learning at the Georgia Institute of Technology and was subsequently expanded well beyond the course scope. The post-course extensions included substantial model refinements, replication experiments across multiple seeds, enhanced data processing, additional analyses, an expanded literature review, and more comprehensive discussions and visualizations. We thank the CS7643 instructional staff for their guidance and constructive feedback during the initial project phase. Computational resources were provided in part by the Georgia Institute of Technology, enabling large-scale training and evaluation of graph neural network models.

{\small
\bibliographystyle{ieee_fullname}
\bibliography{egbib}
}

\section*{A. Project Code Repository}

The GitHub repository for our final project: \href{https://github.com/ffy208/Graph-Neural-Network-for-Product-Recommendation-on-Amazon-Co-purchase-Graph}{link}.

\end{document}